\documentclass[conference]{IEEEtran}
\IEEEoverridecommandlockouts
\usepackage{cite}
\usepackage{amsmath,amssymb,amsfonts}
\usepackage{algorithmic}
\usepackage{graphicx}
\usepackage{textcomp}
\usepackage{adjustbox}
\usepackage{url}
\usepackage{multirow}
\usepackage{booktabs}
\usepackage{xcolor}
\def\BibTeX{{\rm B\kern-.05em{\sc i\kern-.025em b}\kern-.08em
    T\kern-.1667em\lower.7ex\hbox{E}\kern-.125emX}}
\usepackage{fancyhdr}
\title{Non-linear frequency warping using constant-Q transformation for speech emotion recognition}

\usepackage[pscoord]{eso-pic}

\begin{document}

\author{\IEEEauthorblockN{ Premjeet Singh$^1$, Goutam Saha$^2$}
\IEEEauthorblockA{\textit{Dept of Electronics and ECE} \\
\textit{Indian Institute of Technology Kharagpur, Kharagpur, India}\\
premsingh@iitkgp.ac.in$^1$, gsaha@ece.iitkgp.ac.in$^2$}
\and
\IEEEauthorblockN{Md Sahidullah$^3$}
\IEEEauthorblockA{\textit{Universit\'{e} de Lorraine} \\
\textit{CNRS, Inria, LORIA, F-54000, Nancy, France}\\
$^3$md.sahidullah@inria.fr}
}
\maketitle
\thispagestyle{fancy}

\begin{abstract}
In this work, we explore the constant-Q transform (CQT) for speech emotion recognition (SER). The CQT-based time-frequency analysis provides variable spectro-temporal resolution with higher frequency resolution at lower frequencies. Since lower-frequency regions of speech signal contain more emotion-related information than higher-frequency regions, the increased low-frequency resolution of CQT makes it more promising for SER than standard short-time Fourier transform (STFT). We present a comparative analysis of short-term acoustic features based on STFT and CQT for SER with deep neural network (DNN) as a back-end classifier. We optimize different parameters for both features. The CQT-based features outperform the STFT-based spectral features for SER experiments. Further experiments with cross-corpora evaluation demonstrate that the CQT-based systems provide better generalization with out-of-domain training data.
\end{abstract}


\section{Introduction}
The \emph{speech emotion recognition} (SER) is the task for recognizing emotion from human speech. The potential applications of SER include \emph{human-computer interaction}, \emph{sentiment analysis} and \emph{health-care}~\cite{akccay2020speech,krothapalli2013speech,el2011survey, alonso2015new}. Humans naturally sense the emotions in speech while machines find it difficult to characterize them~\cite{picard2000affective, picard2003affective}. Techniques proposed till date have significantly increased the machine's ability to recognize speech emotions. However, the task is still challenging mainly due to the presence of large \emph{interpersonal and intrapersonal variability} and the \emph{differences in speech quality} used to train and evaluate the system. The goal of this work is to develop an improved SER system by considering emotion-specific acoustic parameters from speech that are assumed to be more robust to unwanted variabilities.

Previous studies in SER research have shown that spectral and prosodic characteristics of speech contain emotion-related information. Spectral features include \emph{lower formants frequencies} (F1 and F2), \emph{speech amplitude and energy}, \emph{zero crossing rate} (ZCR) and spectral parameters, e.g., like \emph{spectral flux} and \emph{spectral roll-off}~\cite{chen2012speech,eyben2010towards, eyben2015geneva}. Prosodic features include \emph{pitch}, \emph{pitch harmonics}, \emph{intonation}, and \emph{speaking rate}~\cite{eyben2010towards, eyben2015geneva}. The acoustic front-ends are used with \emph{Gaussian mixture model} (GMM) or \emph{support vector machines} (SVMs) as back-end classifiers for SER tasks~\cite{kockmann2011application}. Studies with prosody reveal that high arousal emotions, such as \emph{Angry}, \emph{Happy} and \emph{Fear}, have higher average pitch values with abrupt pitch variations whereas low arousal emotions like \emph{Sadness} and \emph{Neutral} have lower pitch values with consistent pitch contours~\cite{akccay2020speech,williams1972emotions,banse1996acoustic,cowie1996automatic,deb2018multiscale}. The authors in~\cite{bou2000comparative} have reported that recognition accuracy of Anger is higher near F2 (1250-1750~Hz) and that of Neutral is higher near F1 (around 200-1000~Hz).  Authors in \cite{france2000acoustical} report that center frequencies of F2 and F3 are reduced in depressed individuals. In~\cite{goudbeek2009emotion}, the authors report that high arousal emotions have higher mean F1 and lower F2 and high (positive) valence emotions have high mean F2.  In \cite{bozkurt2011formant}, authors report discrimination between idle and negative emotions using temporal patterns in formants. In~\cite{lech2018amplitude}, the authors have demonstrated that non-linear frequency scales, such as \emph{logarithmic}, mel and \emph{equivalent rectangular bandwidth} (ERB), have considerable impact in SER performance over linear frequency scale.

Recent works with deep learning methods such as \emph{convolutional neural networks} (CNNs) or CNN with \emph{recurrent neural networks} (CNN-RNNs), use spectrogram or raw waveform as input and have shown impressive results~\cite{zhang2017speech,mao2014learning,huang2017characterizing}. These data-driven methods automatically learn the emotion-related representation, however, the role of individual speech attributes in the decision making process is not clear due to the lack of \emph{explainability}. On the other hand, the generalization of these methods remains an open problem, especially when the audio-data for train and test are substantially different in terms of language and speech quality~\cite{parry2019analysis}. 

\begin{figure*}[t]
    \centering
    \includegraphics[scale=0.475]{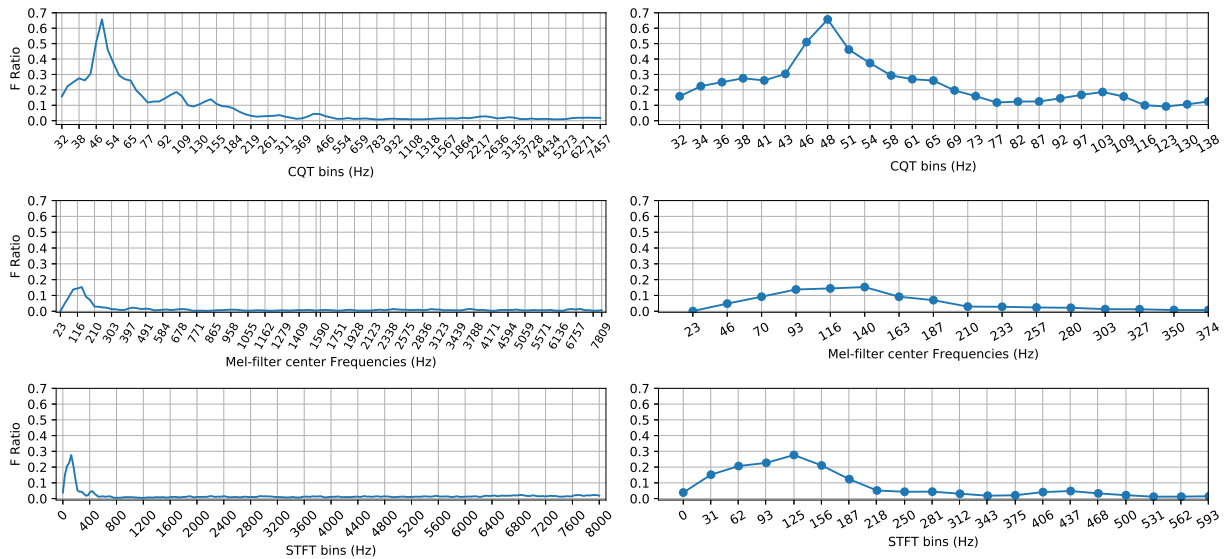}
    \caption{F-ratios of spectrograms based on CQT (top), mel-filter (middle), and standard STFT (bottom) corresponding to the frequency bins. We use the speech sentences with fixed text `a02' from EmoDB database (discussed in Section~\ref{Section:Database}). We select the same text assuming spectra characteristics of emotions to be text-dependent. First column shows the values over the entire frequency range while the second column focuses only on the lower-frequency regions.}
    \label{F_ratio_fig}
\end{figure*}

We address this generalization issue by capturing emotion-related information from speech before processing with a neural network back-end. Given the fact that the low and mid frequency regions of speech spectrum contain pitch harmonics and lower formants that are relevant for emotion recognition, we propose to use a more appropriate approach for time-frequency analysis that produces emotion-oriented speech representation in the first place. Even though the processing with \emph{mel frequency warping} introduces non-linearity in some sense, the power spectrum from the speech is essentially computed with a uniform frequency resolution. We propose to use a time-frequency analysis method called \emph{constant-Q transform} (CQT). This transformation offers higher frequency resolution at low-frequency regions and higher time resolution at high-frequency regions. As the pitch harmonics and lower formants reside in the low-frequency regions of speech spectrum, we hypothesize that keeping high resolution in this region may efficiently capture emotion-related information.

The CQT was initially proposed for music signal processing \cite{brown1991calculation}. Then it was also applied in different speech processing tasks, e.g., \emph{anti-spoofing}~\cite{todisco2017constant, pal2018synthetic}, \emph{speaker verification}~\cite{delgado2016further} and \emph{acoustic scene classification}~\cite{lidy2016cqt}. Recently, the CQT has also been studied for SER~\cite{tang2018end}, but without success. This is possibly due to the lack of optimization of CQT parameters and/or the applied end-to-end model fails to exploit the advantages of CQT. Recent studies show that CNN-based models are suitable for SER including cross-corpora evaluation~\cite{parry2019analysis, issa2020speech}. In this work, we also adopt a CNN-based approach for modeling SER systems. Our main contributions in this work are summarized as follows: (i)~We propose a new framework for CQT-based SER by optimizing CQT extraction parameters for reduced redundancy and improved performance, (ii)~We investigate CNN architecture known as \emph{time-delay neural networks} (TDNNs) suitable for speech pattern classification tasks \cite{21701,snyder2018x} for SER, and (iii)~We perform cross-corpora evaluation with three different speech corpora to assess the generalization ability of the proposed method. Our results demonstrate that the optimized CQT features not only outperform short-time Fourier transform (STFT) features but also provide better generalization.

\section{Methodology}
In this section, we discuss the CQT-based feature extraction framework and the TDNN architecture for emotion recognition.

\subsection{Constant-Q transform}
The CQT of a time-domain signal $x[n]$ is defined as,

\begin{equation}
    X[k]=\frac{1}{N[k]}\sum_{n=0}^{N[k]-1}W[k,n]x[n]e^{-jw_{k}n},
\end{equation}

\noindent where $X[k]$ is the CQT coefficient for $k$-th frequency bin, $W[k,n]$ is the time-domain window for $k$-th bin with duration $N[k]$, $x[n]$ denotes the time samples and $w_{k}=\frac{2 \pi Q}{N[k]}$, where $Q$ is the (constant) Q factor of the filter banks~\cite{brown1991calculation}. In CQT computation, the window length $N[k]$ varies for different values of $k$. Hence, $x[n]$ is correlated with sinusoids of different lengths with equal cycles of oscillation. This leads to constant-Q filter bank representation with geometrically spaced center frequencies over frequency octaves. Hence, we obtain a time-frequency representation which has frequency resolution varying from high to low towards increasing frequencies.

 The CQT representation of an audio signal depends on the \emph{number of octaves of frequencies} and the \emph{number of frequency bins per each octave}. The number of octaves depends upon the chosen minimum frequency ($F_{\mathrm{min}}$) and the maximum frequency ($F_{\mathrm{max}}$) of operation, and this equals to $ \log_{2}{\frac{F_{\mathrm{max}}}{F_{\mathrm{min}}}}$ \cite{todisco2017constant}. The CQT representation with reduced number of total frequency bins over a fixed number of octaves will provide detailed information for lower frequency region with reduced redundancy. Conversely, due to linearly spaced frequency bins, \emph{short-time Fourier transform} (STFT) does not offer this flexibility. Fixing $F_{\mathrm{min}}$ to $32.7$~Hz and $F_{\mathrm{max}}$ to \emph{Nyquist frequency} gives us approximately $8$ octaves. \emph{Hop length} in CQT computation defines the number of time samples by which CQT computation window moves. The CQT also has resemblance with \emph{continuous wavelet transform} (CWT) which provides variable time-frequency resolution and has been found helpful for SER~\cite{shegokar}. 
 
 During the CQT-based feature extraction process, the CQT coefficients are uniformly resampled and then processed with \emph{discrete cosine transform} (DCT) to compute speech features known as \emph{constant-Q cepstral coefficients} (CQCCs).

We perform \emph{class separability} analysis of the time-frequency representations by computing the F-ratios~\cite{nicholson1997evaluating}. The Fig.~\ref{F_ratio_fig} shows the F-ratio obtained at different frequency bins. The higher F-ratios at lower bins for CQT and STFT show the presence of more discriminative information. The figure also indicates that CQT-spectrogram has more number of discriminative coefficients on an average over others due to higher resolution in low-frequency regions.

\subsection{CNN architecture}
The time-frequency representation of speech-like signal is suitable to be used with 1-D CNN, popularly known as TDNN in speech processing literature. Our method is inspired by the TDNN-based \emph{x-vector} system~\cite{snyder2018x} developed for speaker verification task. This processes speech information at \emph{frame} and \emph{segment} level. In frame level, the TDNN captures \emph{contextual information} by applying \emph{kernel} over adjacent frames and by processing each speech frame in an identical manner. This also applies \emph{dilation} in the temporal domain to reduce redundancy and to make it computationally efficient. The frame-level information is processed with several TDNN layers having different kernel sizes and dilation parameters. Finally, \emph{temporal pooling} aggregates frame-level information into segment-level and this is followed by processing with \emph{fully connected} (FC) and \emph{softmax} layer for classification objective. The standard x-vector system computes the segment-level intermediate representation referred as \emph{embeddings} which are further processed with another system for classification. In contrast, our proposed method trains the network in an end-to-end fashion for which the emotion for a test speech is obtained from the output of the trained network.

We empirically optimize the parameters for TDNN architecture. Finally, we use four TDNN layers, followed by statistics pooling with mean and standard deviation, and one FC layer before \emph{softmax}. Table~\ref{tab1} describes the parameters for different layers.  

\begin{table}[ht]
\caption{The parameters of CNN architecture for SER.}
\vspace{0.25cm}
\renewcommand{\arraystretch}{1.2}
\begin{scriptsize}
\begin{tabular}{cccc}
\hline
\textbf{Layer} & \textbf{Size} & \textbf{Kernel Size} & \textbf{Dilation} \\
\hline
TDNN & 32 & 5 & 1 \\ 
TDNN & 32 & 3 & 2 \\
TDNN & 32 & 3 & 3 \\
TDNN & 64 & 1 & 1 \\
Statistics Pooling (Mean and SD)  & 128 & -& - \\
Fully Connected  & 64 & - & - \\
Softmax & \#Classes & - & - \\
\hline
\end{tabular}
\label{tab1}
\end{scriptsize}
\end{table}

\begin{figure*}[ht!]
    \centering
    \hbox{\hspace{1.2cm}\includegraphics[scale=0.4]{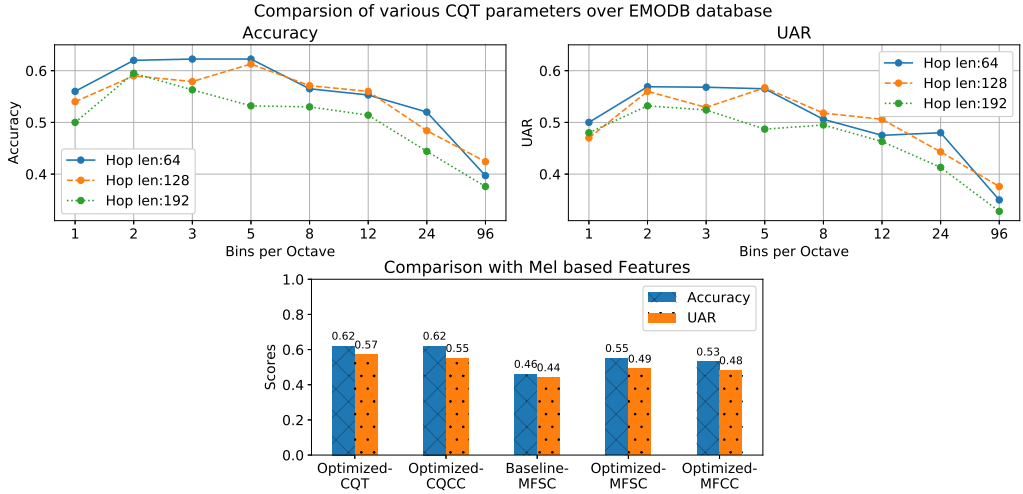}}
    \caption{Performance comparison of CQT for different parameter values. Optimized CQT shows the response of CQT with 3 bins per octave and hop length of 64 samples. Baseline MFSC corresponds to MFSC extraction with standard value of 128 mel-filters and 160 samples hop length, whereas, optimized MFSC has 24 mel-filters with 64 hop. Optimized CQCC and MFCC shown are obtained after applying DCT over optimized MFSC and CQT. The results shown in figure are obtained over EmoDB database only. }
    \label{resfig}
\end{figure*}

\section{Experimental setup}
\subsection{Speech corpora}\label{Section:Database}

In our experiments, we use three different speech corpora which are described in Table~\ref{databasetable}. We downsample speech files at sampling rate of 16~kHz when required. The EmoDB is a German language corpora while RAVDESS and IEMOCAP are in English. For IEMOCAP database, we select only four emotions (Angry, Happy, Sad and Neutral) as some of the emotion class have inadequate data for training neural network models~\cite{issa2020speech}. We perform cross-corpora SER experiments by selecting the same four emotions.

\begin{table}[ht]
\centering
\caption{Summary of the speech corpora used in the experiments. (F=Female, M=Male)}
\vspace{0.25cm}

\begin{scriptsize}
\begin{adjustbox}{width=0.5\textwidth,center}
\label{databasetable}
\begin{tabular}{@{}ccccccc@{}}
\toprule
\textbf{Databases} & \textbf{Speakers} & \textbf{Emotions}  \\ \midrule
\begin{tabular}[c]{@{}c@{}}Berlin Emotion Database \\ (EmoDB) \cite{burkhardt2005database}\end{tabular} & \begin{tabular}[c]{@{}c@{}}10\\ (5~F, 5~M)\\ \end{tabular}  & \begin{tabular}[c]{@{}c@{}}7\\ (Anger, Sad, Boredom, Fear,\\ Happy, Disgust and Neutral)\end{tabular} \\ [0.6cm]

\begin{tabular}[c]{@{}c@{}}Ryerson Audio-Visual Database \\ of Emotional Speech and Song \\(RAVDESS)  \cite{livingstone2018ryerson}\end{tabular} & \begin{tabular}[c]{@{}c@{}}24\\ (12~F, 12~M)\end{tabular} & \begin{tabular}[c]{@{}c@{}}8\\ (Calm, Happy, Sad, Angry, Neutral, \\ Fearful, Surprise, and Disgust)\end{tabular}  \\ [0.6cm]

\begin{tabular}[c]{@{}c@{}}Interactive Emotional Dyadic \\ Motion Capture Database \\ (IEMOCAP) \cite{busso2008iemocap} \end{tabular} & \begin{tabular}[c]{@{}c@{}} 10 \\(5~F, 5~M) \end{tabular} & \begin{tabular}[c]{@{}c@{}} 4 \\ (Happy, Angry, Sad and Neutral) \end{tabular}  \\ [0.6cm]

\bottomrule
\end{tabular}
\end{adjustbox}
\label{datatable}
\end{scriptsize}
\end{table}

\begin{figure*}[ht]
\centering
    \begin{minipage}{0.34\textwidth}
    \includegraphics[width=\textwidth]{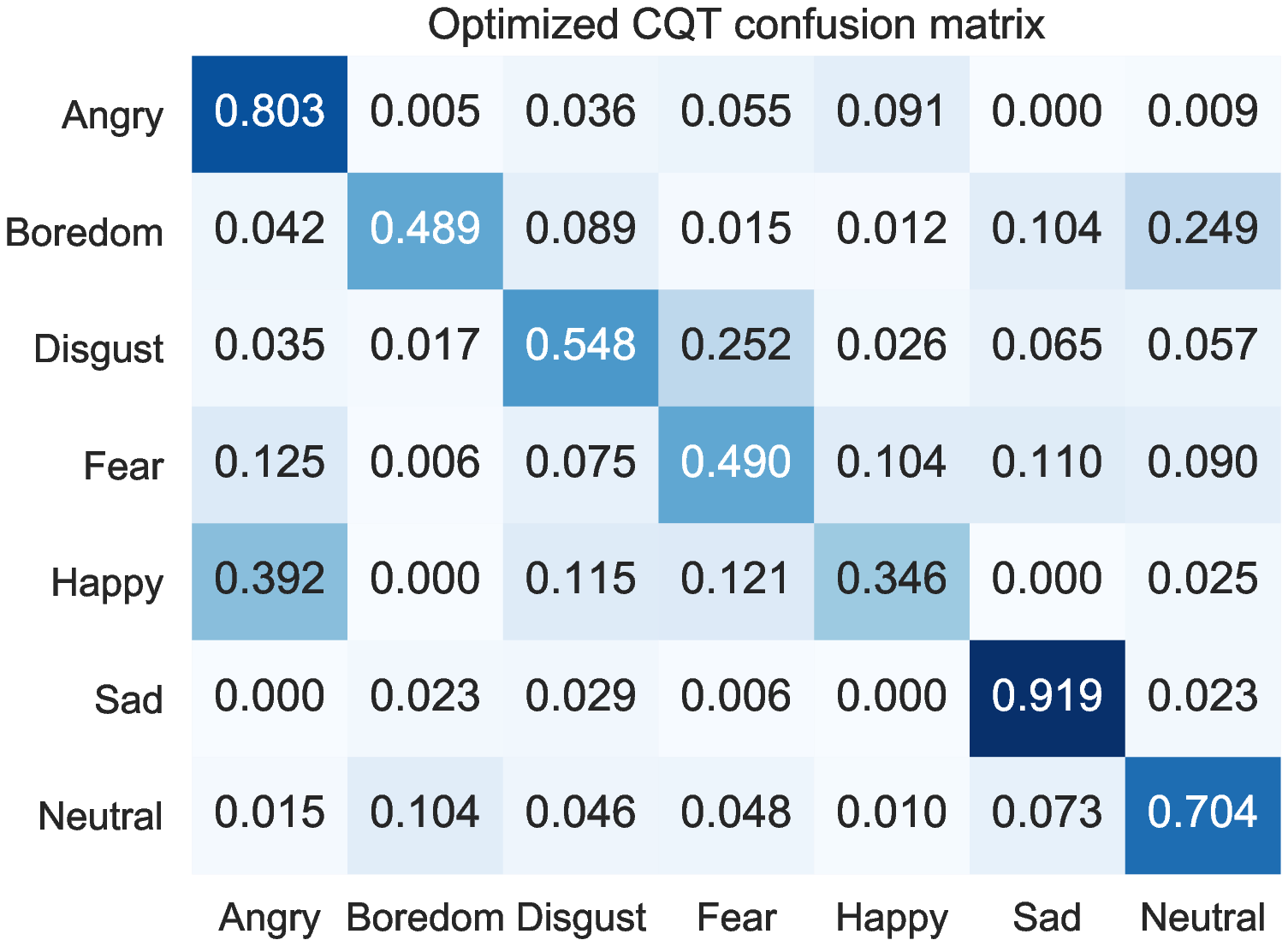}
    \end{minipage}%
    \hspace{2.5cm}
    \begin{minipage}{0.34\textwidth}
    \includegraphics[width=\textwidth]{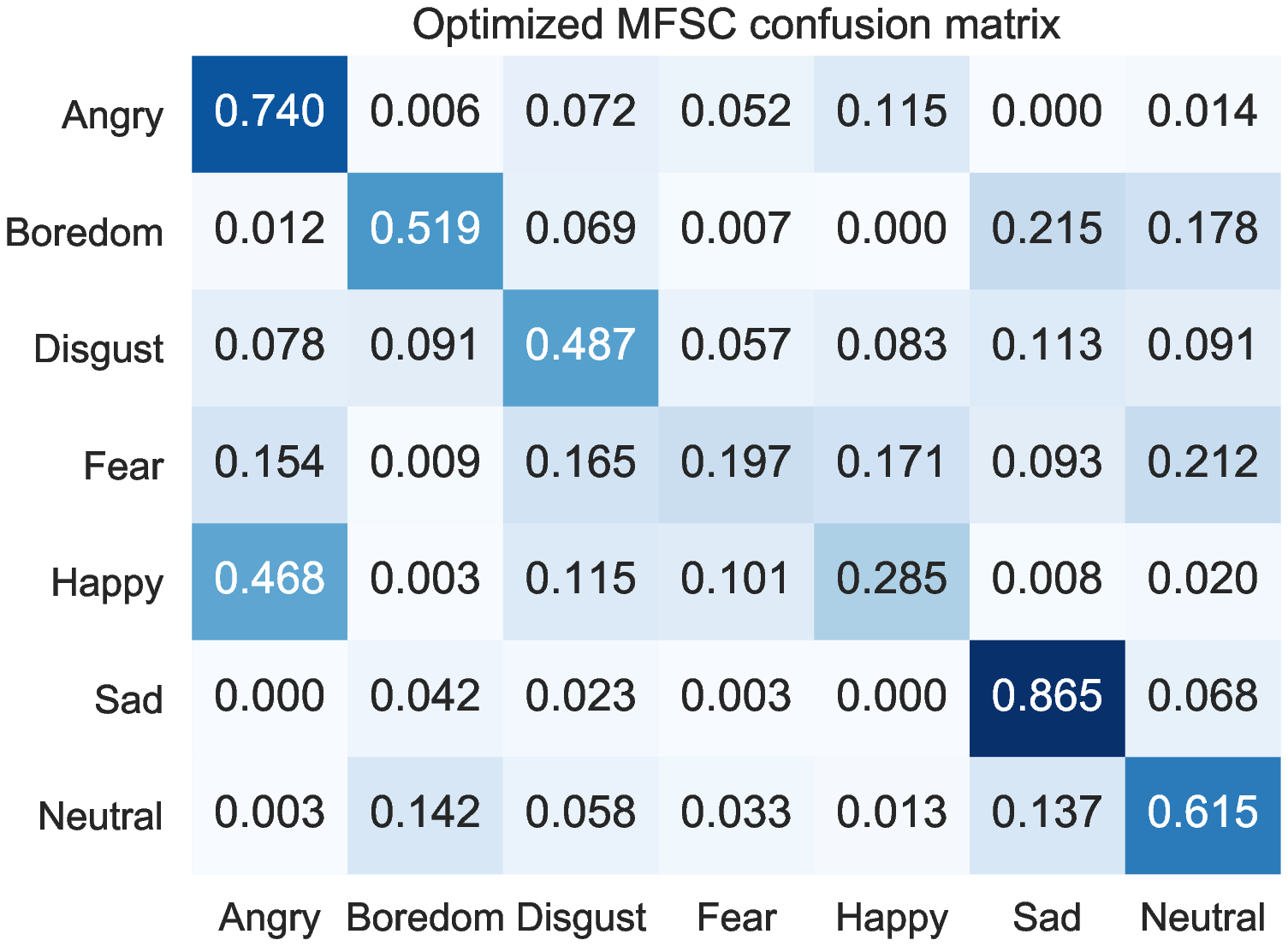}
    \end{minipage}%
    \caption{Confusion matrices for emotion classification experiment with optimized CQT and MFSC features in EmoDB corpus. Given values are the ratio of utterances identified in column class to the total number of utterances in every corresponding row class.}
    \label{Conf_mats}
\end{figure*}

\subsection{Experimental details \& evaluation methodology}

First, we optimize the parameters of the features on EmoDB. We perform experiments on this corpus using \emph{leave-one-speaker-out} (LOSO) cross validation by keeping one speaker in test. Out of the remaining speakers, we use two of them in validation and seven in training. We also apply five-fold data augmentation by corrupting training set with additive noises and room reverberation effect following the \emph{Kaldi} recipe\footnote{\url{https://github.com/kaldi-asr/kaldi/tree/master/egs/voxceleb/v2}} for x-vector training~\cite{snyder2018x}.

We extract features from each speech utterance and discard the non-speech frames with a simple energy-based \emph{speech activity detector} (SAD). We apply utterance-level \emph{cepstral mean variance normalization} (CMVN) before creating the training and validation samples with chunks of $100$ consecutive frames. We consider multiple non-overlapping chunks from the speech utterances depending on the length. We use LibROSA\footnote{\url{https://librosa.github.io/}} python library for feature extraction.

We do not apply chunking for testing and consider the full utterance for computing the test accuracy. We report the final performances with accuracy as well as \emph{unweighted average recall} (UAR). The accuracy is computed as the ratio between the number of correctly classified sentences to the total number of sentences in test. The UAR is given as \cite{rosenberg2012classifying},

\begin{equation}
    \mathrm{UAR} = \frac{1}{K} \sum_{i=1}^{K} \frac{A_{ii}}{\sum_{j=1}^{K} A_{ij}}
\end{equation}

\noindent where $A$ refers to the contingency matrix, $A_{ij}$ corresponds to number of samples in class $i$ classified into class $j$ and $K$ is the total number of classes. As accuracy is considered \emph{unintuitive} for databases with uneven samples across different classes, we optimize the feature extraction parameters based on the UAR metric.

In DNN, we use ReLU activation function and batch normalization for all the hidden layers. For regularization, we apply dropout with probability $0.3$ on the FC layer only. We use Adam optimizer with learning rate $0.001$. The mini-batch size is $64$. We train the models for $50$ epochs and finally testing is done with the model which achieves the highest UAR on the validation set. We repeat each experiment multiple times and report the average performance.

\section{Results}

\subsection{Experiments on EmoDB}
First, we conduct experiments on EmoDB and optimize the CQT parameters. We vary the number of bins per octave from $1$ to $96$. We also perform the experiments with three different hop lengths: $64$, $128$, and $192$. The top row of Fig.~\ref{resfig} shows the standard accuracy and UAR for CQT. We observe improved performance for lower bins per octave and lower hop length. The performance remains very similar for bins per octaves between $2$ and $5$. We select $3$ bins per octave as the optimum observing the consistency in different runs of the experiment. We fix the hop size $64$ as the optimum since the performance is consistently better with this hop size, especially, for lower bins per octave. Since the optimized CQT features use $24$ filters and hop length as $64$, we apply similar configuration for STFT-based \emph{mel frequency cepstral coefficients} (MFCCs) as well as \emph{mel frequency spectral coefficients} (MFSCs) (i.e., MFCCs without DCT). The SER performances with CQT and STFT-based features are illustrated as a bar plot in Fig.~\ref{resfig}. We observe that CQT coefficients as well as CQCCs consistently outperform MFCCs and MFSCs. We also notice that the optimized MFSC outperforms baseline MFSC. The DCT slightly degrades performance in both CQT and STFT-based approaches. We chose the best configuration for both features for the remaining experiments.

\begin{table}[ht]
\renewcommand{\arraystretch}{1.2}
\centering
\caption{Cross-corpora results shown in accuracy / UAR. This uses optimized configuration of MFSC and CQT. To have equal number of classes, only four emotions (Happy, Angry, Sad and Neutral) are considered from every database here. All other parameter settings remain similar to other experiments.}
\vspace{0.25cm}
\label{restablecross}
\begin{tabular}{|c|c|c|c|}
\hline
Train on                   & Test on   & MFSC  & CQT     \\ \hline
\multirow{2}{*}{EmoDB}     & RAVDESS   & 0.41 / 0.44  &   0.44 / 0.46        \\ \cline{2-4} 
                           & IEMOCAP   & 0.36 / 0.37  &   0.38 / 0.39        \\ \hline
\multirow{2}{*}{RAVDESS}   & EmoDB     & 0.45 / 0.42  &   0.48 / 0.48        \\\cline{2-4} 
                           & IEMOCAP   & 0.30 / 0.32  &   0.32 / 0.34        \\\hline
\multirow{2}{*}{IEMOCAP}   & EmoDB     & 0.64 / 0.50  &   0.63 / 0.50          \\ \cline{2-4} 
                           & RAVDESS   & 0.38 / 0.39  &   0.38 / 0.39        \\
                             \hline
\end{tabular}%
\end{table}

Figure~\ref{Conf_mats} shows the confusion matrices obtained for CQT and MFSC in experiments with EmoDB. We observe that CQT is better capable of discriminating emotions such as Fear, Disgust, Sad, Anger and Neutral as compared to MFSC. The CQT-based system yields improved accuracy for Sad, Neutral and Disgust because those emotions are more prominent in low-frequency regions. 
Performance of Boredom is slightly degraded.
Among all the seven emotions, Fear shows the highest gain in performance over MFSC. Happy shows the lowest classification accuracy and a high confusion with Angry.

\subsection{Cross-corpora evaluation}
Table~\ref{restablecross} shows the performance obtained after cross corpus testing. The optimized CQT shows better performance than optimized MFSC for most cases except when the train-test pair are IEMOCAP-EmoDB and IEMOCAP-RAVDESS. The obtained results consolidate our hypothesis that CQT helps in better capturing of emotion-dependent information leading to better generality across databases.

\section{Discussion and Conclusion}
We notice that increasing the frequency resolution at lower frequency regions led to substantial improvement in SER performance. This also confirms that low-frequency region containing pitch harmonics and lower formants convey important emotion-specific information. 
At the same time, the CQT with lower high-frequency resolution does not degrade the overall SER performance which indicates that high-frequency regions are less important from emotion perspective. Also, better performance with fewer frequency bins in both CQT and MFSC indicates less redundant time-frequency representation is more effective for emotion discrimination. Though STFT with optimized parameters generates spectrograms with higher frequency resolution, the performance degrades most likely due to increased redundancy caused by capturing details of high-frequency region. Cross-corpora evaluation suggests that CQT-based time-frequency representation provides better generalization for SER task with different speech corpora in training and test.

We conclude that CQT is a better choice of time-frequency representation in terms of both recognition performance and generalization ability. However, the SER performance is still poor for real-world deployment. We also gain no improvement over MFSC for all the seven emotions included in EmoDB corpus. This indicates that the time-frequency representation needs further investigation for SER. This work can also be extended by exploring CQT representation with recurrent architecture and attention mechanisms which are lacking within our TDNN framework but found useful for SER.

\bibliographystyle{IEEEbib}
\bibliography{mybib}

\end{document}